\documentstyle{article}

\normalsize

\begin{document}

\bibliographystyle{unsrt}
\footskip 1.0cm
\thispagestyle{empty}
\setcounter{page}{0}
\begin{flushright}
IC/95/159\\
July 1995\\
\end{flushright}
\vspace{10mm}

\centerline {\LARGE ON SYMMETRY NON-RESTORATION}
\vspace{5mm}
\centerline {\LARGE AT HIGH TEMPERATURE}
\vspace*{15mm}
\centerline {\large G. Bimonte}
\vspace{5mm}
\centerline{ \small and }
\vspace*{5mm}
\centerline{ \large G. Lozano \footnote{
E-mail addresses: Bimonte@ictp.trieste.it~~,~~Lozano@ictp.trieste.it}}
\vspace*{5mm}
\centerline {\it International Centre for Theoretical Physics, P.O.BOX 586}
\centerline {\it I-34100 Trieste, ITALY}
\vspace*{25mm}
%\baselinestretch{2.0}
\normalsize
\centerline {\bf Abstract}
\vspace*{5mm}
{\large We study the effect of next-to-leading order contributions on the
phenomenon of symmetry non-restoration at high temperature in an
$O(N_1)\times O(N_2)$ model. } \newpage

\baselineskip=24pt
\setcounter{page}{1}

It is a well known fact that broken symmetries are usually
 restored at sufficiently high temperatures \cite{lin,wei,jac}.
Nevertheless, as already noticed
by Weinberg \cite{wei}, there are models, like the $O(N)\times O(N)$
scalar
theory, for which the high temperature phase is less symmetric than the low
temperature one in a certain region of the parameter space.

As noticed by Mohapatra and Senjanovi\'{c} \cite{ms}, the effects of
symmetry non
restoration can have interesting phenomenological implications in
connection to the CP problem. The same phenomenon has also been used to
provide non inflationary solutions to the monopole problem, as
argued by Langacker and Pi \cite{lan} and more recently by Dvali et al.
\cite{dva}.

Weinberg's analysis of  the problem \cite{wei} is based on a lowest order
perturbative calculation of the thermal masses,  the region of parameters
giving
symmetry non restoration being associated to the values of the coupling
constants giving imaginary values for some thermal masses.

In this letter, we will compute the next-to-leading order contributions to the
thermal masses and analyze how the region of symmetry non-restoration
gets modified by their inclusion. The reason for doing this is twofold:
from one side, as these models contain more than one coupling constants,
it may very well happen that next-to-leading order effects dominate over
the leading ones, in certain regions of the parameter space. On the other
side, when trying to achieve symmetry-non restoration in realistic models, such
as
those considered in ref.\cite{dva}, one may have to take values of the
coupling constants large enough for next-to-leading order
corrections to become significant.

For simplicity, we will discuss here only the
simplest model presenting this phenomenon, that is, when
the $O(N_1)\times O(N_2)$ symmetry is global. In this case, the theory
is described by the renormalized euclidean Lagrangian

\begin{equation}
L_R= \frac{1}{2}
\partial_{\mu}\phi_1 \partial_{\mu} \phi_1 + \frac{1}{2}
\partial_{\mu}\phi_2 \partial_{\mu}{\phi_2} + V(\phi_1,\phi_2) + L_{ct}~~,
\label{l1}
\end{equation}
where $\phi_1$( $\phi_2$) is a $N_1$ ($N_2$) components real vector,
$V(\phi_1,\phi_2)$ is a $O(N_1) \times O(N_2)$ symmetric potential:
\begin{equation}
V(\phi_1,\phi_2)= \frac{1}{2} m_1^2 \phi_1 \phi_1+ \frac{1}{2} m_2^2 \phi_2
\phi_2 +
\frac{1}{4!} \lambda_1 \mu^{2\epsilon} (\phi_1 \phi_1)^2 +
\frac{1}{4!} \lambda_2 \mu^{2\epsilon} (\phi_2 \phi_2)^2 +
\frac{1}{4} \lambda \mu^{2\epsilon} (\phi_1 \phi_1)(\phi_2 \phi_2)
\label{pot}
\end{equation}
and $L_{ct}$ stands for the counterterms Lagrangian, calculated in the
$\overline{MS}$ scheme ($\mu$ is the dimensional regularization scale
parameter and $\epsilon=\frac{4-d}{2}$).
As we shall see later, the crucial fact for symmetry non-restoration is that
the mixing coupling constant $\lambda$ can take negative values.
In order for the potential to be bounded from below, the coupling
constants have to satisfy the following relation
\begin{equation}
\lambda_1 \lambda_2 > 9 \lambda^2 ~~.   \label{rel}
\end{equation}

We will show  how next-to leading order corrections to the thermal masses are
calculated,
by first looking at the simpler $O(N)$ case, which can be
obtained from the above Lagrangian by simply setting $\phi_2=0$.
\begin{equation}
L_R= \frac{1}{2}
 \partial_{\mu}\phi_1 \partial_{\mu} \phi_1 +
 \frac{1}{2} m_1^2 \phi_1 \phi_1 +
\frac{1}{4!} \lambda_1 \mu^{2\epsilon} (\phi_1 \phi_1)^2
 + L_{ct}~~.   \label{l2}
\end{equation}

 As
explained in \cite{wei}, perturbation theory based on this Lagrangian breaks
down at high temperatures as powers of the temperature can compensate for
powers of the coupling constants making radiative corrections large. The
remedy to this problem consists in a redefinition  of
the mass and the introduction of a compensating counterterm
 \begin{equation}
L_R= \frac{1}{2}
 \partial_{\mu}\phi_1 \partial_{\mu} \phi_1 +
 \frac{1}{2} M^2 \phi_1^2 +
\frac{1}{4!} \lambda_1 \mu^{2\epsilon} (\phi_1 \phi_1)^2 +
  L_{ct} - \frac{1}{2}\sigma^2 \phi_1 \phi_1 ~~,   \label{l3}
\end{equation}
where
\begin{equation}
M^2=m^2 + \sigma^2
\end{equation}
and $\sigma^2$ is a temperature dependent counterterm (not to be confused
with the mass renormalization counterterm already contained in $L_{ct}$)
which should be determined
self consistently in such a way to eliminate from the self-energy the terms
which diverge
quadratically with the temperature. To illustrate how the mechanism works,
it is
sufficient to consider the case $m^2 \geq 0$. To one loop, one has for the
self energy at zero momentum,
\begin{equation}
\Sigma= -\lambda_1 \frac{2+N_1}{96 \pi^2} M^2
\ln \left(\frac{M^2}{4\pi\mu^2}\right)
-\lambda_1 \frac{2+N_1}{3} T^2 h \left(\frac{M}{T}\right) + \sigma^2~~.
\label{self}
\end{equation}
Here $h$ is defined as:
\begin{equation}
h(y)= \frac{1}{4 \pi^2} \int_0^{\infty} dx
\frac{x^2}{(x^2+y^2)^{\frac{1}{2}} (e^{(x^2+y^2)^{\frac{1}{2}}}-1)}~~.
\label{acca}
\end{equation}
For small values of $y$, $h(y)$ has the asymptotic expansion
\begin{equation}
h(y)= \frac{1}{24} -\frac{1}{8 \pi} y  -\frac{1}{16 \pi^2} y^2
\left(\ln \frac{y}{4\pi} + \gamma - \frac{1}{2}\right)~~,  \label{ase}
\end{equation}
$\gamma$ being the Euler constant.
Notice that in eq.(\ref{self}) we have subtracted  the following UV
divergent part which
is cancelled by the mass counterterm in the $\overline{MS}$ scheme:
\begin{equation}
\Sigma^{UV} = -\lambda_1 M^2 \frac{2N_1+1}{96 \pi^2}[\frac{1}{\epsilon} +
1-\gamma]~.
\end{equation}

 From eq.(\ref{self}), one derives the "gap equation":
\begin{equation}
\frac{M^2}{T^2}= \frac{m^2}{T^2} +
\lambda_1 \frac{ (2+N_1)}{3} h \left(\frac{M}{T}\right)~~,  \label{gap}
\end{equation}
which, after making use of eq.(\ref{ase}), becomes
\begin{equation}
\frac{M^2}{T^2}= \frac{m^2}{T^2} + \lambda_1\frac{ (2+N_1)}{3}
\left(\frac{1}{24}-
\frac{M}{8\pi T}\right) + O (\lambda_1 \frac{M^2}{T^2} \ln\frac{M}{T})~~.
\label{gap2}
\end{equation}
Solving this equation to lowest order gives the well known thermal mass,
\begin{equation}
M^2= m^2 + \frac{2+N_1}{3} \frac{1}{24}\lambda_1 T^2~.   \label{thm}
\end{equation}.

This choice of $M$ would in principle restore the validity of
perturbation theory for $\frac{T}{m} \gg 1$ in the sense that
higher loops corrections will be suppressed by powers of $\lambda$ with
respect to the ``tree level" thermal mass eq.(\ref{thm}). Notice that when
eq.(\ref{thm}) is introduced in eq.(\ref{self}), one gets at high
temperatures ($T \gg m$) \begin{equation}
\Sigma= \frac{3}{\pi}
\left({\frac{2+N_1}{3}} \frac{\lambda_1}{24}\right)^\frac{3}{2} T^2 ~~.
\label{self2}
\end{equation}
This term is the next to leading order correction to the self-energy, and
although it is not immediately obvious, does not get modified by higher
loops corrections, as they are at least of order $\lambda_1^2 \ln
\lambda_1$ (see ref. \cite{par} for details).

Notice that the same contribution could have been obtained more directly
by just solving the gap equation to next to leading order, that is, by
keeping the term linear in $\frac{M}{T}$,
\begin{equation}
\frac{M^2}{T^2}= \frac{m^2}{T^2} + \lambda_1 \frac{(2+N_1)}{3} (\frac{1}{24}-
\frac{M}{8\pi T})~~,   \label{gap3}
\end{equation}
giving, for $\frac{T}{M} \gg 1$
 \begin{equation} M^2=  \frac{2+N_1}{3} \frac{\lambda_1 T^2}{24}
-\frac{3}{\pi}
\left(\frac{2+N_1}{3}\frac{\lambda_1}{24}\right)^{\frac{3}{2}} T^2~~.
\label{gap4}
\end{equation}

At this stage, one would be tempted to improve the result
eq.(\ref{gap4}), by
incorporating also the logarithmic terms in the expansion of $h$.
This would give us a correction to $\frac{M^2}{T^2}$ of order
$\lambda_1^2 \ln \lambda_1$, but, as contributions of this order arise
also from two-loops diagrams, it would be necessary to include them as
well for consistency, something that we will not attempt in this paper.

As we said above, we were assuming that $m^2$ was positive. In fact, the
case in which $m^2$ is negative can be treated in a completely analogous
way, because one is working in the regime of very high temperatures where
one ''self consistently " assumes that the symmetry is restored. By self
consistency we mean that one is able to find solutions to the gap
equations of the unbroken phase giving a positive result for $M^2$.
In this respect, notice that the next to leading correction, although
coming with the opposite sign than the leading one (see eq.(\ref{gap4}),
 will not change
the sign of the self-energy unless $\lambda_1$ becomes large, in which
case any attempt of a perturbative calculation becomes meaningless. Thus,
in this case the subleading correction cannot alter the symmetry breaking
pattern at high $T$.

We now turn to the $O(N_1)\times O(N_2)$ model. There will be a set of two
coupled gap equations,
\begin{eqnarray}
x_1^2 &=& \frac{m_1^2}{T^2} + \lambda_1 \frac{2+N_1}{3} h(x_1) +
\lambda N_2  h(x_2)~~, \label{2gap1}\\
x_2^2 &=& \frac{m_2^2}{T^2} + \lambda_2 \frac{2+N_2}{3} h(x_2) +
\lambda N_1  h(x_1)~~, \label{2gap2} \end{eqnarray}
where
\begin{equation}
x_i=M_i/T
\end{equation}
Using the asymptotic expansion eq.(\ref{ase}), and keeping up to the linear
terms, eqs.(\ref{2gap1}-\ref{2gap2}) become, for $\frac{T}{m} \gg 1$
\begin{eqnarray}
x_1^2 &=& c_1   -\lambda_1 \frac{2+N_1}{24 \pi} x_1
-\lambda \frac{N_2}{8\pi}  x_2 ~~~, \label{lin1}\\
x_2^2 &=& c_2  -\lambda_2\frac{2+N_2}{24 \pi} x_2
-\lambda \frac{N_1}{8\pi}  x_1~~,\label{lin2}
\end{eqnarray}
where we have introduced the constants $c_i$ as
\begin{equation}
c_1 = \lambda_1 \frac{2+N_1}{72}  + \lambda \frac{N_2}{24}~~~,\label{c1}
\end{equation}
\begin{equation}
c_2 = \lambda_2 \frac{2+N_2}{72}  + \lambda \frac{N_1}{24}~~~.\label{c2}
\end{equation}
The constants $c_i$ give the thermal masses $M_i^2$ to lowest order:
\begin{equation}
M_i^2=~c_i~T^2~~.
\end{equation}
As noticed in ref.\cite{wei},
 it is possible to take $\lambda$ negative and still satisfying the
constraint eq.(\ref{rel}), in such a way that one of these combinations of
coupling constants, say $c_2$, is negative (notice that due to the
constraint eq.(\ref{rel}) $c_1$ is necessarily positive), namely:
\begin{equation}
\lambda < 0 \label{reg1a}~~,
\end{equation}
\begin{equation}
\lambda_1 \lambda_2 > 9 \lambda^2 \label{reg1b}~~,
\end{equation}
\begin{equation}
|\lambda|>\frac{2+N_2}{3N_1}\lambda_2~~.\label{reg1c}
\end{equation}
In this case, to lowest order, eqs.(\ref{2gap1}-\ref{2gap2}) have no
self-consistent
real solutions and the corresponding region of parameter space was
identified by Weinberg as the region of symmetry non-restoration (or
rather of symmetry breaking at high $T$, if the symmetry was unbroken at
$T=0$). The question that now arises is whether the subleading
corrections represented by the linear terms in eqs.(\ref{lin1}-\ref{lin2})
can
alter this picture. Notice that, as opposed to the simpler $O(N)$ case,
we now have more than one coupling constants and thus next to leading
order effects could become dominant even when all the couplings are of
the same order of magnitude.
Now, eqs.(\ref{lin1}-\ref{lin2}) represent a pair of parabolae in the plane
$x_1$ and $x_2$ which intersect in the upper
right plane (then giving an acceptable solution to the gap equations and
signalling symmetry restoration at high $T$) if and only if:
\begin{equation}
8\pi \frac{c_2}{\lambda N_1} \leq
\left\{ -\lambda_1 \frac{2+N_1}{48 \pi} +
\left[ \left(\lambda_1\frac{2+N_1}{48 \pi}\right)^2 +
c_1\right]^\frac{1}{2} \right\}~~.\label{reg2}
\end{equation}
The corresponding region of symmetry non-restoration is then described by the
set of inequalities:
\begin{equation}
\lambda <0~~\label{reg3a},
\end{equation}
\begin{equation}
\lambda_1 \lambda_2 > 9 \lambda^2~~\label{reg3b},
\end{equation}
\begin{equation}
|\lambda|> \lambda_2 \frac{2+N_2}{3N_1}+|\lambda| \frac{3}{\pi}
\left\{ -\lambda_1 \frac{2+N_1}{48 \pi} +
\left[ \left(\lambda_1\frac{2+N_1}{48 \pi}\right)^2 +
c_1\right]^\frac{1}{2} \right\}~~.\label{reg3c}
\end{equation}
By comparing with eqs.(\ref{reg1a}-\ref{reg1c}), it is clear that
the inclusion of
next-to-leading order corrections reduces the region of the parameter space
in which symmetry non-restoration takes place.
We observe that even if there is always a
region of symmetry non restoration, its size becomes smaller and smaller
as $N_1$ and $N_2$ grow large.

We can think of our model as representing the Higgs sector of a $SU(5)$
grand unified model, like the one considered in ref.\cite{dva}. In
this case, taking $N_1=90$, $\phi_1$ represents the real components
of the Higgs in the ${\bf 45}$ representation of the group $SU(5)$,
while taking
$N_2=24$, $\phi_2$ corresponds to its adjoint ${\bf 24}$ representation.

A qualitative indication of the importance of the subleading terms can
be obtained by plotting the regions (\ref{reg1a}-\ref{reg1c}),
(\ref{reg3a}-\ref{reg3c}) at fixed values
of one of the coupling constants and of $N_1$ and $N_2$. Figures (1) and
(2) display these regions for $\lambda_1=4/5$ and $\lambda_2=1$
respectively and $N_1=90$, $N_2=24$, which correspond to typycal
values of the $SU(5)$ parameters considered in ref.\cite{dva}. It is
clear from both figures that the subleading terms have a non negligible
effect, as they reduce (in the considered region of parametrs) by a factor
of roughly two the size of the region of symmetry non-restoration.

In conclusion, we have shown that the inclusion of next-to-leading order
contributions can modify in a substantial way the symmetry breaking
pattern of an $O(N_1) \times O(N_2)$ model at large $T$.
Naturally, the phenomenologically interesting models are those in which part
of this symmetry is gauged. Consequently, our analysis
of the gap equations should be generalized to include this case. The
results presented here lead us to expect that subleading corrections will play
an important role also in grand-unified models. We expect to report on this
issue in a forthcoming publication.

{\bf Acknowledgements}

We would like to thank G.Senjanovi\'{c} and Alejandra Melfo for useful
discussions and for communicating us their results prior to publication of
their work.

\newpage

\centerline {\LARGE FIGURE CAPTIONS}
\vspace{30mm}

Figure 1. [Region of symmetry non-restoration at fixed $\lambda_1=4/5$ for
$N_1=90$ and $N_2=24$. The zeroth order region is the one enclosed by the
solid curves, while the region obtained by including next-to-leading order
corrections is the one enclosed by the upper solid line and the dashed line.]

\vspace{15mm}

Figure 2. [Region of symmetry non-restoration at fixed $\lambda_2=1$ for
$N_1=90$ and $N_2=24$. The zeroth order region is the one enclosed by the
solid curves, while the region obtained by including next-to-leading order
corrections is the one enclosed by the upper solid line and the dashed line.]

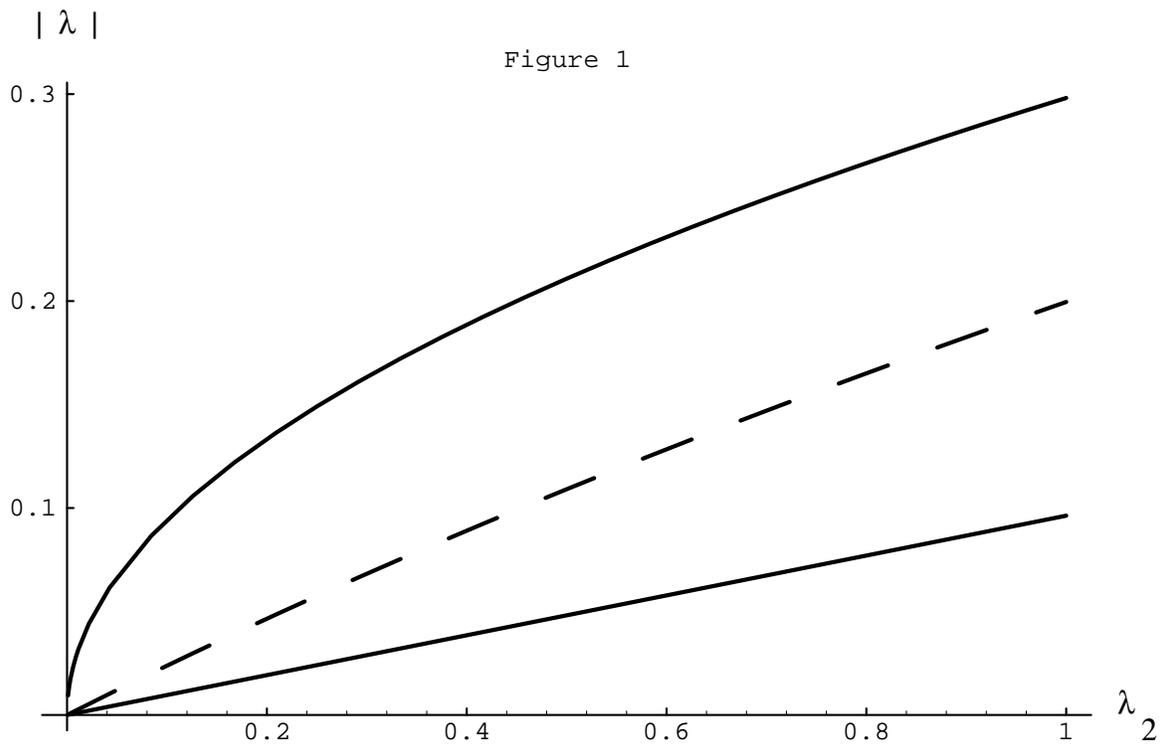
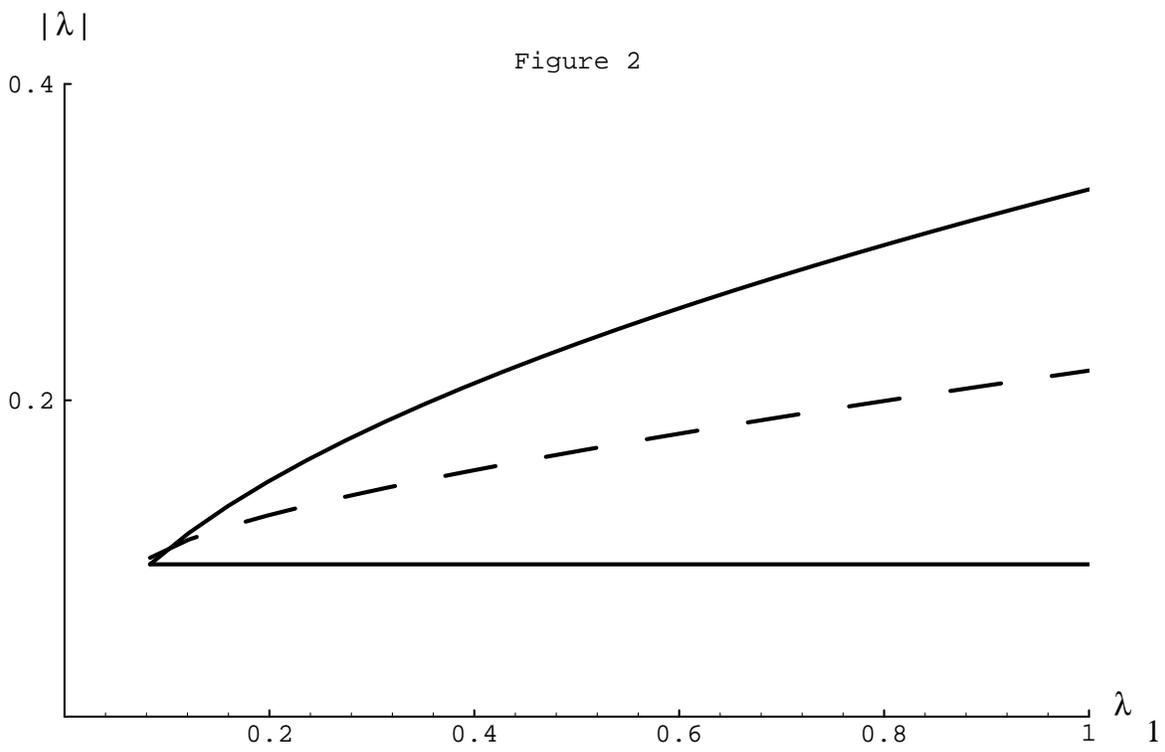
\end{document}